\documentclass[prb,showpacs,twocolumn]{revtex4}%

\usepackage{amsfonts}
\usepackage{amsmath}
\usepackage{amssymb}
\usepackage{graphicx}%
\begin{document}
\title{Mean-field potential calculations of high-pressure equation of state for
shock-compressed BeO}
\author{Qili Zhang, Ping Zhang, Haifeng Song, Haifeng Liu}
\affiliation{Institute of Applied Physics and Computational Mathematics, P.O. Box 8009,
Beijing 100088, P.R. China}

\begin{abstract}
A systematic study of the Hugoniot equation of state, phase transition, and
the other thermodynamic properties including the Hugoniot temperature, the
electronic and ionic heat capacities, and the Gr\"{u}neisen parameter for
shock-compressed BeO, is presented by calculating the total free energy. The
method of calculations combines first-principles treatment for 0-K and
finite-T electronic contribution and the mean-field-potential approach for the
vibrational contribution of the lattice ion to the total energy. Our
calculated Hugoniot shows good agreement with the experimental data.

\end{abstract}
\maketitle

As one member of the series of alkaline-earth oxide, Beryllium oxide (BeO) has
many unique properties. First, it is the only one alkaline-earth oxide to
crystallize in the wurtzite (WZ) rather than the rock-salt (RS) structure.
Further, not only is BeO harder than the other alkaline-earth oxides but also
it is among the hardest materials known. It is a good insulator like other
alkaline-earth oxide, but its heat conductivity is an order of magnitude
higher, which makes it a technically promising ceramic\cite{Vidal}. Also, BeO
has a high melting point. These interesting physical properties are related to
characteristic features of interatomic bonding in BeO: for example, Compton
scattering measurements revealed a significant component of the primarily
ionic bonding in BeO\cite{Joshi}.

An important issue for BeO is its pressure induced phase transition sequence,
which has been investigated by a few experimental\cite{Hazen,Jeph} and
theoretical studies\cite{Chang1,Chang2,Lich1,Lich2,Boe}. This is motivated by
the well-known dielectric theory of Phillips and Van Vechten\cite{Phillips}
that the structural phase stabilities of the $A^{N}B^{8-N}$ compounds can be
related to their ambient spectroscopic ionicities ($f_{i}$). According to this
theory, binary compounds with $f_{i}>0.785$ crystallize in a sixfold
coordinated structure such as RS, while those with $f_{i}<0.785$ crystallize
in a fourfold structure such as zincblende (ZB) or wurtzite (WZ). BeO has a
Phillips ionicity\cite{Phillips} of 0.602 and crystallize in the WZ structure
under ambient conditions, in accord with the dielectric theory. Under
pressure, the dielectric theory\cite{Phillips} predicts that the tetragonal
compounds with $f_{i}>0.35$ first transform to an ionic sixfold structure and
then to some metallic phase, while those with $f_{i}<0.35$ transforms to the
metallic phase directly. Given this understanding, however, the pressure
induced phase sequence for BeO is not yet fully understood. The early two
\textit{ab initio }calculations support this general prediction that a
WZ$\rightarrow$RS transition occurs at 22 GPa \cite{Chang1} (40 GPa from
Ref.\cite{Jeph}). On the other side, the latter calculations predicted the WZ
first transformed into zinc blende (ZB) and then into RS. Herein, the work of
Camp and Doren using soft nonlocal pseudopotential predicted the WZ-ZB-RS
transitions at 74 and 137GPa\cite{Camp}, while Park \textit{et al.}%
\cite{Park}, who used a first-principles soft nonlocal pseudopotential method
within the generalized-gradient approximation (GGA), predicted the transition
pressure to be 91 (WZ$\rightarrow$ZB) and 147 GPa (ZB$\rightarrow$RS)
respectively. This unusual transition sequence is attributed to the large
charge asymmetry and the small bond length of BeO\cite{Park}. By use of
full-potential linear muffin-tin orbital (FPLMTO) method, Boettger and
Wills\cite{Boe} have also predicted a WZ$\rightarrow$ZB$\rightarrow$RS phase
transition sequence with the final transition at about 95 GPa. They also
predicted a theoretical Hugoniot, consistent with shock wave data up to 100
GPa. More recently, Cai \textit{et al.}\cite{Cai}, by systematically
calculating the enthalpy barrier of the phase transition, have shown that only
the WZ$\rightarrow$RS transition will occur with increasing pressure.

Experimentally, Raman spectra showed no evidence for a phase transition up to
55 GPa in BeO\cite{Jeph} and Hugoniot data did not reveal any volume change
for stresses up to 100 GPa\cite{Jeph}. The most recent static high pressure
x-ray diffraction experiment\cite{Yoshi} also showed no phase transition up to
126 GPa, but a new phase was found at 137 GPa. With increasing pressure, the
WZ phase disappeared at 175 GPa.\cite{Yoshi}

In this paper we theoretically study the phase transformation and Hugonoit of
BeO under shock-wave dynamic compression, which is one of the most efficient
ways to explore the thermodynamic properties of a material at ultrahigh
pressures and temperatures. From its derived Hugoniot state one can deduce
much useful information such as equation of state (EOS) at feasible
temperature range. The theoretical methodology we employed in this paper is a
combined \textit{ab initio} electronic structure calculation and classical
mean-field potential (MFP) thermodynamic treatment\cite{WangYi}. Recent
efforts have applied the MFP approach to a number of prototype
metals\cite{WangYi}, indicating that both the calculated Hugoniot equation of
state (at pressures up to $\sim$1000 GPa and temperatures up to $\sim$70 000
K) and room temperature isotherms are well described within the experimental
uncertainties. We calculate the 0K enthalpy of WZ, ZB, and RS BeO as a
function of pressure. The results show that the enthalpy of WZ BeO is equal to
that of ZB BeO at 84 GPa. When pressure is increased to be 105 GPa, the
enthalpy for WZ and RS BeO approach to cross. Based on our first-principles
total energy data, we give a MFP analysis of the phase transition, Hugoniot
EOS and the other thermodynamic properties for shock-compressed BeO. Our
calculated Gibbs free energies show that the WZ$\rightarrow$RS trnasition
pressure is 103.8 GPa. Furthermore, our calculated Hugoniot is in good
agreement with the experimental data.

For a system with a given averaged atomic volume $V$ and temperature $T$, the
Helmholtz free-energy $F(V,T)$ per atom can be written as
\begin{equation}
F(V,T)=E_{c}(V)+F_{ion}(V,T)+F_{el}(V,T), \tag{1}%
\end{equation}
where $E_{c}$ represents the 0-K total energy which is obtained from
\textit{ab initio} electronic structure calculations, $F_{el\text{ }}$is the
free energy due to the thermal excitation of electrons, and $F_{ion}$ is the
ionic vibrational free energy$\ $which is evaluated from the partition
function $Z_{ion}=\exp(-NF_{ion}/k_{B}T)$. Here $N$ is the total number of
lattice ions. In the mean-field approximation, the classical $Z_{ion}$ is
given by\cite{Wasserman}
\begin{equation}
Z_{ion}=\left(  \frac{mk_{B}T}{2\pi\hbar^{2}}\right)  ^{3N/2}\left(  \int
\exp\left(  -g(\mathbf{r},V\right)  /k_{B}T)d\mathbf{r}\right)  ^{N}. \tag{2}%
\end{equation}
The essential of the MFP approach is that the mean-field potential
$g(\mathbf{r},V)$ is simply constructed in terms of $E_{c}$ as
follows\cite{WangYi}
\begin{equation}
g(r,V)=\frac{1}{2}\left[  E_{c}(R+r)+E_{c}(R-r)-2E_{c}(R)\right]  , \tag{3}%
\end{equation}
where $r$ represents the distance that the lattice ion deviates from its
equilibrium position $R$. It should be mentioned that the well-known
Dugdale-MacDonald expression\cite{Dug} for the Gr\"{u}neisen parameter can be
derived by expanding $g(r,V)$ to order $r^{2}$. Then, $F_{ion}$ can be
formulated as
\begin{equation}
F_{ion}(V,T)=-k_{B}T\left(  \frac{3}{2}\ln\frac{mk_{B}T}{2\pi\hbar^{2}}+\ln
v_{f}(V,T)\right)  , \tag{4}%
\end{equation}
with
\begin{equation}
v_{f}(V,T)=4\pi%
{\displaystyle\int}
\exp\left(  -\frac{g(r,V)}{k_{B}T}\right)  r^{2}dr. \tag{5}%
\end{equation}

When the electron-phonon interaction and the magnetic contribution are
neglected, the electronic contribution to the free energy is $F_{el}%
=E_{el}-TS_{el}$, where the bare electronic entropy $S_{el}$ takes the
form\cite{Jarlborg}
\begin{equation}
S_{el}(V,T)=-k_{B}%
{\displaystyle\int}
n(\epsilon,V)\left[  f\ln f+(1-f)\ln(1-f)\right]  d\epsilon, \tag{6}\label{e1}%
\end{equation}
where $n(\epsilon,V)$ is the electronic density of states (DOS) and $f$ is the
Fermi distribution. With respect to Eq. (6), the energy $E_{el}$ due to
electron excitations can be expressed as
\begin{equation}
E_{el}(V,T)=\int n(\epsilon,V)f\epsilon d\epsilon-\int^{\epsilon_{F}%
}n(\epsilon,V)\epsilon d\epsilon, \tag{7}%
\end{equation}
where $\epsilon_{F}$ is the Fermi energy. Given the Helmholtz free-energy
$F(V,T)$, the other thermodynamic functions such as the entropy $S=-(\partial
F/\partial T)_{V}$, the internal energy $E=F+TS$, the pressure $P=-(\partial
F/\partial V)_{T}$, and the Gibbs free energy $G=F+PV$, can be readily calculated.

The first task is to calculate the 0-K total energy $E_{c}(V)$ in a wide range
of the volume. For this we employ the full-potential
linearized-augmented-plane-wave (FPLAPW) method\cite{Blaha} within the
generalized gradient approximation (GGA)\cite{Perdew}. The total energy of BeO
with WZ structure was calculated at 96 lattice constants ranging from 4.14
a.u. to 6.04 a.u. in steps of 0.02 a.u., while for the ZB structure 91 lattice
constants ranging from 6.22 a.u. to 8.02 a.u. in steps of 0.02 a.u. were
considered. Finally for RS BeO we sample 81 lattice constants ranging from
5.46 a.u. to 8.66 a.u. in steps of 0.04 a.u.. As a reference, we mention that
the calculated zero-pressure lattice constants for the three BeO structures
are $a=5.135$ a.u. ($c/a=1.623$, WZ), $7.232$ a.u. (ZB), and $6.894$ a.u.
(RS), respectively. In all of three structures of BeO and for all the atomic
volumes considered, we use the constant value of the muffin-tin radii $R_{mt}$
of 1.2 a.u. for Be atom and 1.0 a.u. for O atom respectively. The plane-wave
cutoff $K_{cut}$ is determined by $R_{mt}\times K_{cut}=10.0$. 4000 $k$-points
in the full Brillouin zone are used for reciprocal-space integerations.%

\begin{figure}[tbp]
\begin{center}
\includegraphics[width=1.0\linewidth]{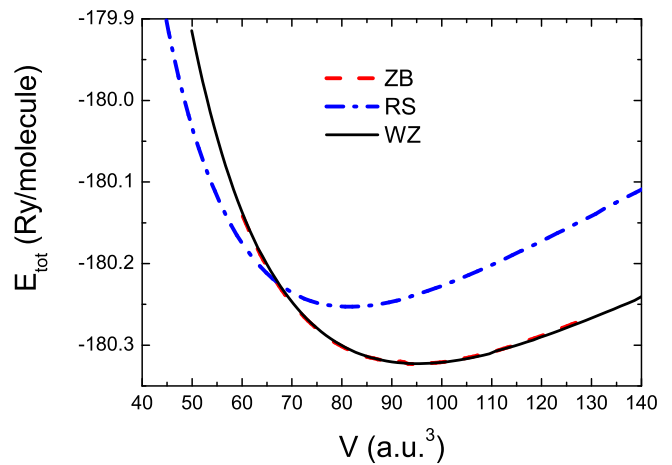}
\end{center}
\caption{(color online). Calculated total energies for WZ,ZB,and RS phase
of BeO are plotted as a function of volume.
}
\label{fig1}
\end{figure}%
Figure 1 shows the calculated 0-K total energies as a function of volume for
WZ, ZB, and RS phases of BeO. The equilibrium volumes (per BeO molecule)
corresponding to the minima in total energy are $95.53$ a.u. $^{3}$ (WZ),
$95.02$ a.u. $^{3}$ (ZB), and $82.18$ a.u. $^{3}$ (RS), respectively. The
equilibrium properties, including lattice parameters, bulk modulus from the
fit with Murnaghan curve\cite{Mur1}, and the equilibrium elastic constants for
the three crystal structures of BeO, are listed in Table I. For comparison,
the available experimental data are also given in Table I. The other available
theoretical data have been collected in Ref.[2]. For WZ phase, the calculated
equilibrium lattice constant and bulk modulus are 2.712 \r{A}and 203 GPa, in
good agreement with the experimental values of 2.698 \r{A}and 212
GPa\cite{Hazen}. Also one can see that the calculated values of elastic
constants $c_{ij}$ for WZ structure agree as well with the measured data.

\begin{table}[th]
\caption{Calculated and experimental structural parameters for WZ, ZB, and RS
phases of BeO: equilibrium lattice constants $a$ (\AA ), bulk modulus $B_{0}$
(GPa), and the elastic constants $c_{ij}$ (GPa). Note that the $c/a$ ratio for
the WZ has been fully optimized.}%
\label{table4}
\begin{tabular}
[c]{ccccccccccc}\hline\hline
Phase & $a$ & $c/a$ & $B_{0}$ & $c_{11}$ & $c_{12}$ & $c_{13}$ & $c_{33}$ &
$c_{44}$ & $c_{66}$ & \\\hline
WZ & 2.712 & 1.623 & 203 & 432 & 120 & 87 & 463 & 142 & 156 & \\
WZ (exp.$^{a}$) & 2.698 & 1.622 & 212 & 460 & 125 & 82 & 490 & 145 & 167 & \\
ZB & 3.825 & - & 201 & 342 & 148 & - & - & 208 & - & \\
RS & 3.648 & - & 231 & 298 & 214 & - & - & 294 & - & \\\hline\hline
$^{a}$Reference 23. &  &  &  &  &  &  &  &  &  & \\
&  &  &  &  &  &  &  &  &  &
\end{tabular}
\end{table}%

\begin{figure}[tbp]
\begin{center}
\includegraphics[width=1.0\linewidth]{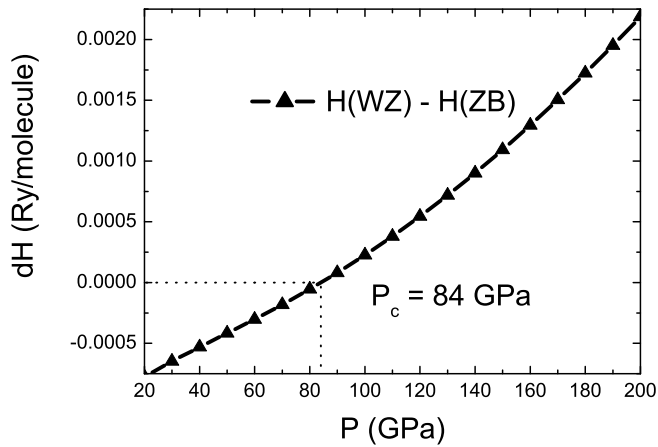}
\end{center}
\caption{Calculated enthalpy difference between WZ and ZB type BeO
at 0 K as a function of pressure.
}
\label{fig2}
\end{figure}%
To obtain the most stable structure at finite pressure and temperature, the
Gibbs free energy $G=E+PV-TS$ should be used. For the experimental data, the
last term is assumed small and therefore usually neglected by the previous
theoretical studies\cite{Chang1,Chang2,Lich1,Lich2,Boe}. For comparison with
those calculations, we also work with the enthalpy $H=E+PV$ for each phase as
a first step. The enthalpy of BeO with WZ, ZB, and RS structures are
calculated, and the phase transition pressure is obtained from the enthalpy
curve crossings. The difference in enthalpy between WZ and ZB type BeO as a
function of pressure is plotted in Fig. 2, which shows that the two structures
are energetically very close with the enthalpy difference typically of $\sim$5
meV in a wide range of pressure. This result is not surprising because of the
high similarity between the two structures. The local environment of any atom
in either ZB or WZ is exactly the same as far as the second neighbors. One can
see from Fig. 2 that the enthalpy of WZ BeO is lower than that of ZB phase up
to 84 GPa of pressure, while the enthalpy of ZB BeO begins to be lower for
$P>84$ GPa. It thus seems that the zero temperature WZ$\rightarrow$ZB
transition pressure is 84 GPa. This result is close to that (87 GPa) obtained
by Cai \textit{et al}.\cite{Cai} based on pseudopotential method with
generalized-gradient approximation (GGA). From the point of view of enthalpy
barrier, however, Cai \textit{et al}. noticed that the WZ$\rightarrow$ZB phase
transition cannot happen up to 200 GPa. Here we adopt this point and will not
consider the WZ$\rightarrow$ZB transition in the following Hugoniot calculations.%

\begin{figure}[tbp]
\begin{center}
\includegraphics[width=1.0\linewidth]{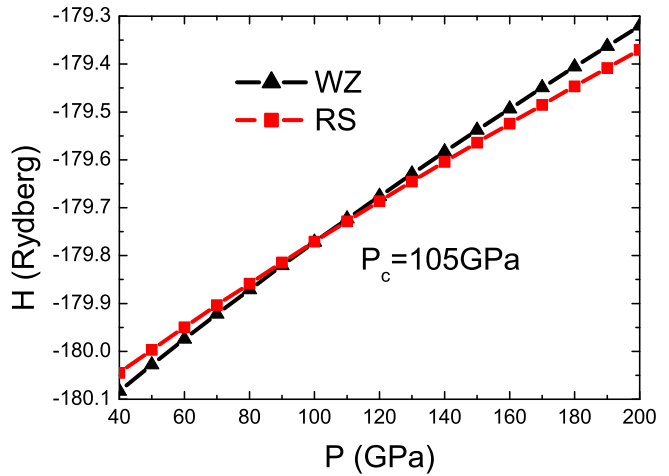}
\end{center}
\caption
{(color online). Calculated enthalpy of WZ and RS type BeO at 0 K as a function of pressure.
}
\label{fig3}
\end{figure}%
The enthalpy of BeO with WZ and RS structures are shown in Fig. 3 as a
function of pressure. One can see that the enthalpy of the two structures
increase almost linearly with increasing pressure. When the pressure is up to
105 GPa they become equal. With further increasing pressure the RS structure
becomes more stable. Thus it is expected to undergo a WZ$\rightarrow$RS phase
transition around $P$=105 GPa. This value is identical to that obtained in
Ref.\cite{Cai} but prominently differs from the calculated result of 147 GPa
in Ref.\cite{Park}. This difference seems to be mainly caused by the different
methods for finding the transition pressure.

Next we investigate the phase transition properties of BeO under shock-wave
compress. In this case, we fully take into account the temperature effect in
MFP framework and calculate the Gibbs free energies $G$ of the Hugoniot states
for the WZ and RS structures, respectively.
\begin{figure}[tbp]
\begin{center}
\includegraphics[width=1.0\linewidth]{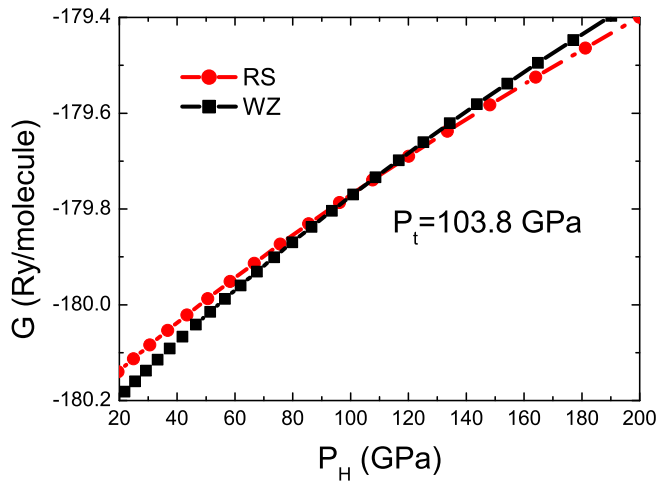}
\end{center}
\caption
{(color online). Calculated Gibbs free energies of the WZ and RS type BeO
as a function of the Hugoniot pressure
}
\label{fig4}
\end{figure}%
Fig. 4 shows the Gibbs free energies of these two structures as a function of
the Hugoniot pressure. It can be seen that the overall slope and dependence of
Gibbs free energy upon pressure for both two structures are essentially the
same as zero-temperature case [Fig. 3]. The Hugoniot WZ$\rightarrow$RS
transition pressure is 103.8 GPa. Compared to zero-temperature result (105
GPa), one can see that the temperature effect (typically of 1000 K, see Fig. 7
below) is small for BeO WZ$\rightarrow$RS phase transition. Thus we derive a
conclusion that the enthalpy can be reasonably used to study the phase
transition of BeO in experimentally relevant temperature region. This is not
surprising considering the fact of high melting point of BeO.

Now we turn to study the Hugoniot for BeO. Hugoniot states, which are derived
by the conventional shock-wave technique\cite{Los}, are characterized by using
measurements of shock velocity ($U_{s}$) and particle velocity ($U_{p}$) with
$V_{H}/V_{0}=(U_{s}-U_{p})/U_{s}$ and $P_{H}=\rho_{0}U_{s}U_{p}$, where
$P_{H}$ is the Hugoniot pressure, and $\rho_{0}$ is the initial density. By
applying the Rankine-Hugoniot jump conditions, these data define a compression
curve [volume ($V_{H}$) versus pressure ($P_{H}$)] as a function of known
Hugoniot energy ($E_{H}$):
\begin{equation}
P_{H}(V_{0}-V)=2(E_{H}-E_{0}) \tag{8}%
\end{equation}
where $V_{0}$ and $E_{0}$ refer to the atomic volume and energy under ambient
condition, respectively.%

\begin{figure}[tbp]
\begin{center}
\includegraphics[width=1.0\linewidth]{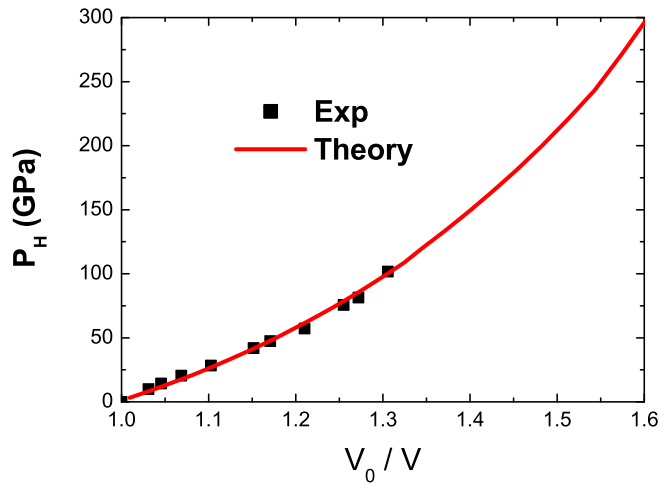}
\end{center}
\caption
{(color online). Calculated Hugoniot (solid line) in the form of Hugoniot pressure
versus $V_{0}/V$ plot and experimental data (squares).
}
\label{fig5}
\end{figure}%
Unlike the static EOS, the temperature along the Hugoniot can undergo a range
from room temperature to several tens of thousands of degrees, thus the
calculations of the Hugoniot state could serve as a good check of a
theoretical method for the thermodynamic calculation. Using the calculated
$V_{0}$ and $E_{0}$, we calculate the Hugoniot EOS for BeO at pressures up to
200 GPa. Due to the fact revealed in Fig. 4 that there occurs a WZ$\rightarrow
$RS phase transition at the Hugoniot pressure $P_{t}=103.8$ GPa, thus during
calculation of Hugoniot EOS, we choose BeO structure to be WZ for the Hugoniot
pressure $P_{H}<P_{t}$, while for $P_{H}>P_{t}$ the RS phase of BeO is chosen.
Correspondingly, the initial ambient volume ($V_{0}$) and energy ($E_{0}$) of
WZ phase is used when $P_{H}<P_{t}$, while for $P_{H}>P_{t}$, the values of
$V_{0}$ and $E_{0}$ refer to the RS phase. Our calculated Hugoniot EOS is
shown in Fig. 5 (solid curve) in the form of a Hugoniot pressure ($P_{H}$)
versus $V_{0}/V$. Note that to reproduce the experimental density\cite{Hazen},
in accord with the commonly used expedient, here the 0-K isotherm of WZ
structure was uniformly adjusted by reducing 5.85 GPa to each pressure and the
energies are subsequently modified in a consistent fashion. For comparison
with our theoretical result, the experimental data\cite{Los} are also shown in
the figure (squares). One can see that an overall agreement between our
calculation and the experiment is fairly good up to the experimentally
attainable Hugoniot pressure of 102 GPa, except for four points in the range
of 0-30 GPa where the strength effects are large. At present, unfortunately,
no experimental shock-wave data at further high compression ($P_{H}>102$ GPa)
exist. Therefore, to verify our theoretical Hugoniot for RS phase at
$P_{H}>P_{t}$ GPa, more experimental data are needed. However, based on the
good agreement between our calculations and the experiment for WZ phase BeO,
and the coincidence between our calculated result of WZ-RS phase transition
and the previous calculation\cite{Cai}, we expect that our calculated Hugoniot
curve of $P_{H}$ versus $V_{0}/V$ can be adopted in the realistic application
if the higher-pressure data are needed.%

\begin{figure}[tbp]
\begin{center}
\includegraphics[width=1.0\linewidth]{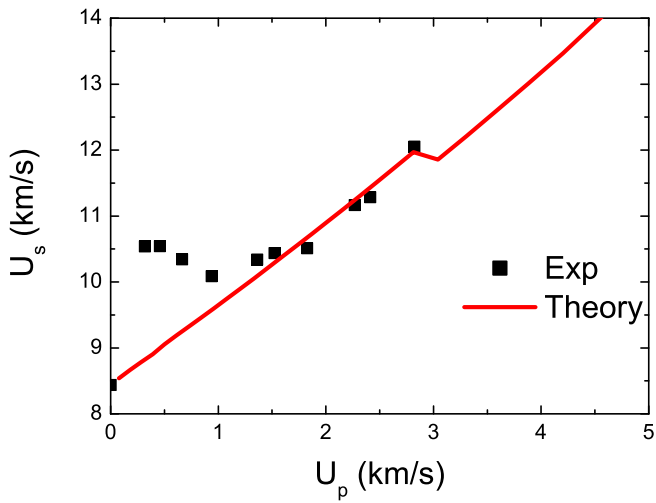}
\end{center}
\caption{(color online). Calculated Hugoniot in the form of $U_{s}%
$ versus $U_{p}$
plot (solid line) and experimental data (squares).
}
\label{fig6}
\end{figure}%
For further illustration, we have also transformed the calculated $P_{H}$
versus $(V_{0}/V)$ relation to that between shock velocity $U_{s}$ and
particle velocity $U_{p}$. The result is plotted in Fig. 6 (solid curve). The
experimental data\cite{Hazen} are also shown in this figure (squares) for
comparison. One can see that at mediate particle velocities (2 km/s$<U_{p}<$3
km/s), the calculation agrees well with the experiment. Whereas at low
particle velocities the discrepancy becomes apparent although the calculated
starting point ($U_{p}=0$) coincides exactly with the measurement (see Fig.
6). Boettger \textit{et al}. have also noticed this discrepancy in their
zero-temperature calculation\cite{Boe}. This may be due to the experimental
failure to correctly measure the particle velocity $U_{p}$ at low
shock-pressure region by the presence of strength effect as mentioned above.%

\begin{figure}[tbp]
\begin{center}
\includegraphics[width=1.0\linewidth]{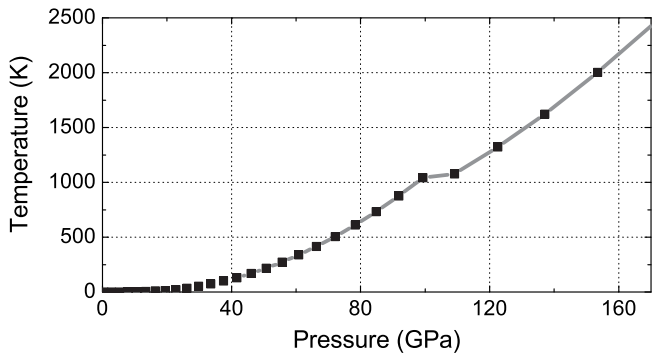}
\end{center}
\caption{Calculated temperature along the principal Hugoniot for BeO.
During calculation the crystal phase of BeO has been chosen to be
wurtzite at pressures lower than WZ-RS transition point $P_{t}$, and
rocksalt for $P>P_{t}$.
}
\label{fig7}
\end{figure}%
We also carried out the calculation of the temperature along the principal
Hugoniot. Note that in the traditional reduction of the Hugoniot data, the
temperature estimate remains less secure since it requires the accurate
knowledge of specific heat and the Gr\"{u}neisen parameter values that are not
well known. Using the present MFP, however, all these quantities can be
calculated straightforwardly (see below). The calculated Hgoniot emperature is
shown in Fig. 7 as a function of pressure. During calculation, again, we
choose BeO structure to be WZ at Hugoniot pressure $P_{H}<P_{t}$, while for
$P_{H}<P_{t}$ the RS phase of BeO is chosen. The turning segment in Fig. 7
around $P_{t}$ is due to this choice of different crystal phases in the two
pressure regions separated by transition point $P_{H}=P_{t}$. At present there
is no experimental data of Hugonoit temperature available. Thus our calculated
result needs to be verified in the future.%

\begin{figure}[tbp]
\begin{center}
\includegraphics[width=1.0\linewidth]{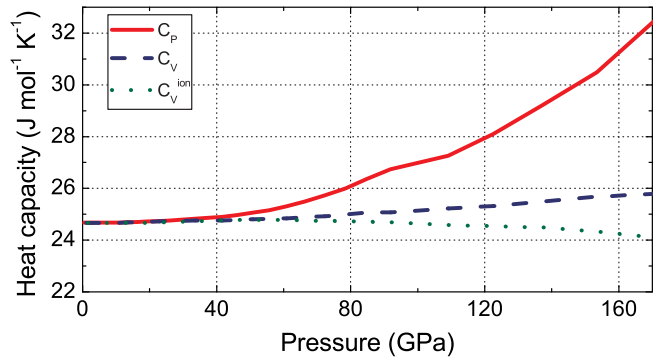}
\end{center}
\caption{Calculated thermodynamic Gr\"{u}%
neisen parameter along the principal Hugoniot for BeO.
During calculation the crystal phase of BeO has been chosen to be
wurtzite at pressures lower than WZ-RS transition point $P_{t}$, and
rocksalt for $P>P_{t}$.
}
\label{fig9}
\end{figure}%

Since we have explicitly calculated the Hermholtz free energy $F(V,T)$ as a
function of $V$ and $T$, all other thermodynamic parameters can be calculated.
In particular, one can evaluate the thermodynamic Gr\"{u}neisen parameter.
Given the isothermal bulk modulus $B_{T}$ which is given by $B_{T}%
(V,T)=V\left(  \frac{\partial^{2}F}{\partial V^{2}}\right)  _{T}$, the volume
thermal-expansion coefficient $\beta_{P}(V,T)=\frac{1}{V}\left(
\frac{\partial V(T)}{\partial T}\right)  _{P}$, and the constant-volume heat
capacity $C_{V}=C_{V}^{\text{ion}}(V,T)+\left(  \frac{\partial E_{el}%
(V,T)}{\partial T}\right)  $, the thermodynamic Gr\"{u}neisen parameter can be
obtained as follows:
\begin{equation}
\gamma_{\text{th}}=\frac{VB_{T}(V,T)\beta_{P}(V,T)}{C_{V}(V,T)}.
\tag{9}\label{E9}%
\end{equation}
Figure 8 plots our calculated $\gamma_{\text{th}}V_{0}/V$ as a function of
Hugoniot pressure $P_{H}$. Again two crystal structures, i.e., the wurtzite
phase for $P_{H}<P_{t}$ and the rocksalt phase for $P_{H}<P_{t}$, are used in
the calculation. One can see that the conventional assumption $\gamma
/V$=constant for the reductions of shock-wave data is not applicable on the
whole for the present case of BeO.%

\begin{figure}[tbp]
\begin{center}
\includegraphics[width=1.0\linewidth]{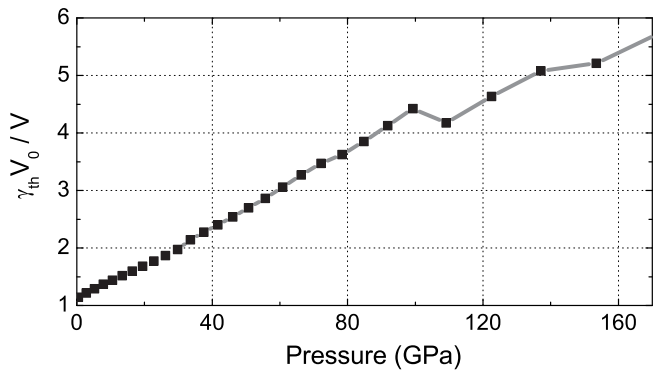}
\end{center}
\caption
{(color online). Calculated heat capacities along the principal Hugoniot for BeO.
During calculation the crystal phase of BeO has been chosen to be
wurtzite at pressures lower than WZ-RS transition point $P_{t}$, and
rocksalt for $P>P_{t}%
$. The solid, dashed, and dotted curves represent the calculated constant-pressure specific heat
$C_{P}$, constant-volume specific heat $C_{V}$, and the lattice ion only
constant-volume $C_{V}^{ion}$, respectively.
}
\label{fig8}
\end{figure}%
Figure 9 plots the constant-pressure heat capacity $C_{P}$ which is calculated
by
\begin{equation}
C_{P}(V,T)=C_{V}(V,T)+VTB_{T}(V,T)\beta_{P}^{2}(V,T).\tag{10}\label{E10}%
\end{equation}
For comparison, the constant-volume heat capacity $C_{V}$ and its ionic
component $C_{V}^{\text{ion}}$ are also shown in Fig. 9. One can see that the
thermal electronic contribution (the difference between the dashed line and
dotted line) to the heat capacity is negligibly small in a wide range of
pressure. This is what one expects considering the insulator nature of BeO.
For the metals, on the other side, the thermal electronic contribution to
$C_{V}$ may be comparable with the lattice ion contribution at high
compression due to the increasing electronic DOS at fermi level.

In summary, by using first-principles FPLAPW total-energy calculation method
supplemented with MFP\ treatment to take into account the vibrational
contribution of the lattice ions, we have systematically studied the Hugoniot
EOS of shock-compressed BeO. Our calculated Hugoniot shows good agreement with
the experimental data. The other thermodynamic properties, such as the
Hugoniot temperature, the Gr\"{u}neisen parameter, and the heat capacities
including the ionic and electronic contributions, have also been calculated.
In addition, the 0-K enthalpy and finite-temperature Gibbs free energies of
BeO with WZ, ZB, and RS structures have also been calculated, which indicates
that the WZ$\rightarrow$RS transition pressure change by Hugoniot temperature
is tiny. Therefore, the 0-K calculation is considerably reasonable in studying
the solid-solid phase transition of BeO. We expect our present results shed
some light on understanding the Hugoniot properties of oxides under shock-wave compression.

This work was partially supported by CNSF under grant number 10544004 and 10604010.

\end{document}